\documentclass[usegraphicx,usenatbib]{mn2e}
\usepackage{txfonts}
\usepackage{graphicx}
\usepackage[Symbol]{upgreek}
\usepackage[T1]{fontenc}

\def\degr{\hbox{$^\circ$}}
\def\arcmin{\hbox{$^\prime$}}
\def\farcs{\hbox{$.\!\!^{\prime\prime}$}}

\title[Double--double radio galaxy J1706$+$4340]{Multifrequency study of
a double--double radio galaxy J1706$+$4340}
\author[A. Marecki, M. Jamrozy \& J. Machalski]
{A. Marecki$^{1}$\thanks{E-mail: \texttt{amr@astro.uni.torun.pl}},
M. Jamrozy$^{2}$, J. Machalski$^{2}$\\
$^{1}$Toru\'n Centre for Astronomy,
Faculty of Physics, Astronomy and Informatics,
Nicolaus Copernicus University,
Grudzi\k{a}dzka 5, PL-87-100 Toru\'n, Poland\\
$^{2}$Astronomical Observatory, Jagiellonian University,
ul. Orla 171, PL-30-244 Krak\'ow, Poland}

\begin{document}

\date{Accepted 2016 August 8. Received 2016 August 5; in original form 2016 June 20}

\pagerange{\pageref{firstpage}--\pageref{lastpage}} \pubyear{0000}

\maketitle

\label{firstpage}

\begin{abstract} 

We report the outcome of multifrequency radio observations of a double--double
radio source J1706$+$4340 carried out with the Very Large Array and Giant
Metrewave Radio Telescope. After supplementing our own data with those
available in the literature, we collected a considerable set of radio
measurements covering the range from 74 to 8460\,MHz. This has enabled us to
perform a comprehensive review of physical properties of the source and its
dynamical evolution analysis. In particular, we found that, while the age of
the large-scale outer lobes is in the range $260\!-\!300$\,Myr, the renewal
of the jet activity, which is directly responsible for the double--double 
structure, took place only about 12\,Myr ago after about 27-Myr long period 
of quiescence. Another important property of J1706$+$4340 we found is
that the injection spectral indices and the jet powers for the inner and the
outer doubles are very similar. This implies that it is the spin of the
supermassive black hole rather than e.g. an instability of the accretion disc
that is likely responsible for the jet production and its properties.

\end{abstract}


\section{Introduction}

Lobes of radio sources associated with active galactic nuclei (AGNs) are huge 
reservoirs of energy so even when fuelling provided by the jets is cut-off,
they are still observable, albeit in a relic form, for a substantial amount 
of time. \cite{Komissarov1994} argue that the so-called coasting phase of 
the lobes may last up to $10^8$\,yr hence it is quite likely that 
ignition of a new active episode can still take place during the coasting 
phase, i.e. before the lobes have faded completely. The observable effect of 
this coincidence would be the presence of a new pair of radio lobes 
straddled with an old, larger, double-lobed relic structure. If both pairs 
of lobes are co-linear and if there is a common centre of symmetry located 
at the galactic nucleus then the whole such structure is labelled a 
double--double radio source (DDRS). The list of best-known DDRSs and the 
review of their properties has been given by \cite{Saikia2009}.

Since the outer lobes of DDRSs are no longer fuelled, they have steep 
spectra due to ageing, thus they are best observable at low radio 
frequencies. To attain high resolution at that part of the spectrum, 
numerous DDRSs have been targeted by Giant Metrewave Radio Telescope (GMRT). 
Nevertheless, the most comprehensive approach in such investigations is to 
combine the GMRT low-frequency data with the Very Large Array (VLA) data 
obtained at higher frequencies. To date, only few DDRSs, e.g. J1453$+$3308 
(\citealt{Konar2006}) and PKS\,1545$-$321 (\citealt{Machalski2010}), have 
been investigated this way. Owing to determination of their spectra over a 
wide frequency range, many of the astrophysical parameters could be 
established, including the ages of both inner and outer lobes as well 
as the length of the quiescence period between two epochs of activity 
the two pairs of lobes are signatures of.

In this article, radio data for J1706$+$4340 gathered in the range from 74 
to 8460\,MHz, predominantly by means of the GMRT and VLA observations, are 
analysed in order to study the dynamics and energetics of this source. The 
paper is organized as follows. The discovery of the J1706$+$4340 as a DDRS 
is introduced in Section\,2, optical and infrared data of the host galaxy 
are quoted in Section\,3, and multifrequency radio data are presented in 
Section\,4. The dynamical analysis of each pair of radio lobes of 
J1706$+$4340 is elaborated upon in Section\,5. The results are discussed in 
Section\,6 and summarized in Section\,7.

The cosmological parameters by
\cite[][$H_0\!=\!71\,{\rm km\,s}^{-1}{\rm Mpc}^{-1}, \Upomega_{\rm M}\!=\!0.27$,
and $\Upomega_{\Uplambda}\!=\!0.73$]{Spergel2003} are used throughout this
article. Positions are given in the J2000.0 coordinate system.

\begin{table*}
\caption{Radio data of J1706$+$4340 analysed in this study}
\begin{tabular}{r c l r r r c c}
\hline
Frequency  & Telescope/         &Date of        & \multicolumn{3}{c}{Beam size and PA}                  & rms       &Ref. \\
(MHz)      & Survey             &observation    & ($^{\prime\prime}$) & ($^{\prime\prime}$) & ($^\circ$) &(mJy beam$^{-1}$)&      \\
           &                    &               &                     &                     &            &           &      \\ 
(1)        &    (2)             &    (3)        &  \multicolumn{3}{c}{(4)}                               & (5)       &  (6) \\ 
\hline
           &            	&            	&                     &                     &            &           &      \\
  73.8     &VLA-B/VLSSr 	& 2003 Sep  20	&   75                & 75                  &            &  180.7    &  1   \\
 151.5     &Cambridge/6C	& 1976--1978  	&   252.0             & 364.9               &            &   40      &  2   \\   
 152.2     &GMRT        	& 2011 Aug  25 	&   24.78             & 21.33               &$-$36.60    &  5.19     &  p   \\
 232.0     &MSRT        	& 1985--1993   	&   228.0             & 330.2               &            &  50.0     &  3   \\            
 240.1     &GMRT        	& 2011 Aug  31 	&   36.85             & 10.63               &$-$69.20    &  2.92     &  p   \\
 326.9     &WSRT/WENSS  	& 1991        	&   54                & 78.2                &            &  3.25     &  4   \\
 408.0     &Northern Cross/B3	& 1977 Feb--May 	&   156               & 288.96              &            &  10       &  5   \\
 612.2     &GMRT        	& 2011 Aug  31	&   5.82              & 4.82                & $-$3.36    &   0.071   &  p   \\     
1400.0     &VLA-D/NVSS  	& 1995 Mar  12	&   45                & 45                  &            &   0.45    &  6   \\ 
1400.0     &VLA-B/FIRST 	& 1994 Aug  19  &   5.4               & 5.4                 &           &   0.18    &  7   \\ 
1425.0     &VLA-A       	& 2006 Feb  21  &   1.35              & 1.06                &29.99       &   0.0764  &  p   \\
1425.0     &VLA-B       	& 2005 Apr  15 	&   4.13              & 3.64                &$-$20.27    &   0.099   &  p   \\
4860.1     &VLA-B       	& 2005 Apr  15  &   1.21              & 1.07                &$-$27.24    &   0.053   &  p   \\
4860.1     &VLA-C             	& 2005 Aug  9   &   3.89              & 3.56                &$-$25.00    &   0.059   &  p   \\
4850.0     &Green Bank 91m/GB6	& 1986--1987  	&   222               & 194                 &            &   5       &  8   \\ 
8460.1     &VLA-B             	& 2005 Apr  15  &   0.71              & 0.64                &$-$34.99    &   0.034   &  p   \\  
8460.1     &VLA-C             	& 2005 Aug  9   &   2.29              & 2.10                &$-$50.32    &   0.041   &  p   \\
\hline
\end{tabular}
\begin{flushleft}
References along with the names of some of the well-known surveys.
(1) VLSSr: \cite{Lane2014};
(2) 6C: \cite{Hales1988};
(3) MSRT: \cite{Zhang1997}; 
(p) this paper; 
(4) WENSS: \cite{Rengelink1997}; 
(5) B3: \cite{Ficarra1985};
(6) NVSS: \cite{Condon1998}; 
(7) FIRST: \cite{Becker1995}; 
(8) GB6: \cite{Gregory1996}\\
\end{flushleft}
\end{table*}

\section{The double--double nature of J1706$+$4340}

J1706$+$4340 (B3\,1704$+$437) has not been studied in detail to date
although its double--double nature has been recognized. \cite{Proctor2011} 
listed J1706$+$4340 as one of 242 double--doubles identified in Faint Images 
of the Radio Sky at Twenty-cm survey at 1400\,MHz (FIRST; \citealt{Becker1995}).
That list has been critically reviewed by \cite{Nandi2012} who confirmed the 
double--double nature of only 23 objects including J1706$+$4340. However, this 
study was begun in 2004, triggered by the work presented by \cite{Marecki2006}
also related to restarted activity in AGNs but focused on the so-called
core-dominated triple (CDT) sources. CDTs could be the early stages of DDRSs
when the inner double is still too compact to be resolved in the maps
encompassing the whole double--double structure.

\cite{Marecki2006} devised an automated procedure to select CDTs from FIRST.
Each FIRST source with a flux density listed in that catalogue being greater
than a particular threshold (75\,mJy) was picked and then a pair of sources
straddling it, i.e. the potential lobes, was sought within a 2-arcmin radius
of the initially selected source -- the potential core. It was required that
the peak flux densities of the putative lobes were less than 30 per cent of
that of the putative core. That threshold was somewhat arbitrary, but was
based upon an initial, extensive visual inspection of FIRST maps: the peak
flux densities of the lobes in the vast majority of sources with the required
morphology fell well below that limit anyway.

Owing to the above selection procedure, it was possible to select many CDTs 
but it had one side effect that was not destructive to the CDT selection
process and in the end turned out to be fortunate. If a potential CDT core
was accompanied with another nearby compact source and if such a double shared
a common pair of lobes, the algorithm formally detected two CDT candidates,
each centred on one of the two point-like sources but sharing the same pair
of the outer lobes. It follows that identifying such cases was equivalent to
a selection of potential DDRSs. In order to make sure that those were indeed
DDRSs, the positions of optical objects near the centre of the field were
taken into account. Eventually, it was found that in one such case neither
of the two alleged CDT cores was an actual one but both straddled the
optical/infrared object. Because of the presence of the common pair of
large-scale lobes found by the algorithm, the whole selection process by
\cite{Marecki2006} led to a serendipitous discovery of a new DDRS --
J1706$+$4340 -- in which the point-like sources appear as the inner pair of
mini-lobes co-linear with the outer pair of extended but diffuse lobes.

\begin{table*}
\caption{Flux densities of different components of J1706$+$4340}
\begin{tabular}{r c r r r r r}
\hline
Frequency &\multicolumn{6}{c}{Flux density (mJy)}       \\
(MHz)    &\multicolumn{1}{c}{Total structure} & \multicolumn{3}{c}{$<$----------- Inner structure ----------$>$}& \multicolumn{2}{c}{Outer structure}\\  
         &  Observed          &  Observed     &   Observed     &{\sl Model fit}& Observed&{\sl Model fit} \\
         &                    &NE lobe  &SW lobe  &NE+SW \\
\hspace{2mm}(1) & (2)         &    (3)        &      (4)       &    (5)     &   (6)        &  (7) \\
\hline
         &                    &               &                &            &              &            \\
73.8     & 1419$\pm380^{a}$   &               &                &            &              &            \\
151.5    & 1170$\pm$110       &               &                &            & 708.8$\pm$110&{\sl 692.0} \\
152.2    & 1125$\pm$60        &               &                &            & 665.7$\pm$60 &{\sl 688.8} \\  
232.0    &  650$\pm65^{b}$    &               &                &            & 323.5$\pm$65 &{\sl 453.7} \\  
240.1    &  878$\pm$47        &               &                &            & 560.3$\pm$47 &{\sl 438.3} \\ 
326.9    &  548$\pm$29        &               &                &            & 300.9$\pm$30 &{\sl 320.1} \\
408.0    &  429$\pm20^{c}$    &               &                &            & 222.3$\pm$20 &{\sl 254.2} \\
612.2    &  334$\pm$17        &20.19$\pm$1.04 &134.44$\pm$6.73 &{\sl 147.3} & 179.6$\pm$16 &{\sl 164.3} \\     
1400.0   &  148$\pm$8         &               &                & {\sl 78.2} &\\
1400.0   &                    & 9.02$\pm$0.72 & 70.66$\pm$3.59 & {\sl 78.2} & 68.2$\pm$7.5 &{\sl  60.8} \\
1425.0   &                    & 7.08$\pm$0.38 & 62.01$\pm$3.14 & {\sl 77.1}      \\ 
1425.0   &                    & 9.36$\pm$0.54 & 66.13$\pm$3.32 & {\sl 77.1}      \\
4850.0   & 36.1$\pm$7.4       &               &                &            &  7.4$\pm$3.6 &{\sl 8.20} \\
4860.1   &                    & 3.27$\pm$0.21 & 25.03$\pm$1.30 & {\sl 27.6}      \\
4860.1   &                    & 3.26$\pm$0.20 & 25.85$\pm$1.27 & {\sl 27.6}      \\
8460.1   &                    & 2.05$\pm$0.13 & 15.04$\pm$0.76 & {\sl 16.8}      \\
8460.1   &                    & 2.23$\pm$0.14 & 16.28$\pm$0.87 & {\sl 16.8}      \\
\hline
\end{tabular}
\begin{flushleft}
$^a$The original flux density of the VLSSr survey (RBC scale; 
\citealt{Roger1973}) was multiplied by a factor of 0.9 to suit the 
\cite{Baars1977} scale;\\
$^b$this is the peak flux and the error is 10 per cent of the flux;\\
$^c$the original B3 flux density which is given in the CKL scale
(\citealt{Conway1963}) is according to \cite{Baars1977} multiplied by 
a factor of 1.129 to the common Baars scale.\\
\end{flushleft}
\end{table*}

\section{Optical and Infrared Data}
\label{opt_data}

According to Sloan Digital Sky Survey (SDSS; \citealt{Adelman2008}), the 
position of the stellar-like object associated with J1706$+$4340 is 
RA=17$\rm^{h}$06$\rm^{m}$25\fs44, Dec.=+43\degr40\arcmin40\farcs16. Its SDSS 
optical magnitudes are: u=$22.03\!\pm\!0.16$, g=$20.79\!\pm\!0.03$, 
r=$19.81\!\pm\!0.02$, i=$19.26\!\pm\!0.01$, and z=$18.79\!\pm\!0.03$. Its
photometric redshift based on the sixth SDSS release is
$z\!=\!0.525\!\pm\!0.051$. This makes 1\,arcsec on the sky to be an
equivalent of 6.242\,kpc.

The {\em Wide-field Infrared Survey Explorer} satellite
(\citealt{Wright2010}) catalogue gives the following magnitudes of 
J170625.43+434040.1: W1($3.4\mu$m)=$14.676\!\pm\!0.027$, 
W2($4.6\mu$m)=$13.957\!\pm\!0.031$, W3($12\mu$m)=$11.415\!\pm\!0.135$, and 
W4($22\mu$m)=$8.557$. According to \cite{Gurkan2014} and \cite{Wright2010}, 
the calculated colours (W1-W2)=0.719 and (W2-W3)=2.542 indicate that the 
central source is of an AGN type and that the parent object is located close 
to the area occupied by quasars. Using the infrared photometric data given 
above as well as the Two Micron All-Sky Survey (\citealt{Skrutskie2006}) 
measurements, we fitted the spectral energy distribution (SED) of the parent 
object with the Code Investigating GALaxy Emission ({\sc cigale)} 
package\footnote{{\sc cigale} code is publicly available at 
http://cigale.lam.fr.} (\citealt{Roehlly2014}). The SED is concordant with 
the SDSS magnitudes and the resultant photometric redshift is slightly 
smaller but -- accounting for its uncertainty -- still consistent with the 
SDSS estimate given above.

\section{Multifrequency radio observations}

For the purpose of this study, we analysed the data we acquired ourselves 
using the VLA and GMRT but also those available in the archives. Details of
the observations are shown in Table\,1 and in the three following subsections.
The flux densities of different parts of the radio structure of J1706$+$4340
are listed in Table\,2.

\subsection{GMRT observations and data analysis}
\label{GMRT_data}

We performed dedicated observations of J1706$+$4340 with GMRT in three 
frequency bands: 152, 240, and 612\,MHz, the latter two being obtained in dual 
frequency mode. The project code was 20$_-$005 and the observations were 
carried out on 2011\,August 25 and 31 for the 152 and 240/612\,MHz bands, 
respectively. Data were recorded using 8-s integration time with the 
available frequency band divided into 256 channels. The usual scheme of 
observing phase calibrator interlaced with the observation of the target 
source was adopted. Flux density calibrators 3C286 and 3C48, respectively, 
were observed for about 15\,min. at the beginning and at end of each 
observing run. Phase calibrators 3C309.1 and 3C286 were used at frequencies 
centred at 152 and 240/612\,MHz, respectively. The total integration 
time on the target source was about 5.3 and 5.5\,h at 152 and 240/612\,MHz,
respectively.

Data reduction was carried out following standard calibration and reduction
procedures in Astronomical Image Processing System ({\sc aips}). Data were
edited for strong radio frequency interference and then standard flux, phase,
and bandpass calibrations were applied. The deconvolved images were made using
the task {\tt IMAGR}, where the field of view covering the primary beam was
subdivided into a number of facets. Several rounds of phase-based
self-calibration were performed. The resultant image was then corrected for
the primary beam using the task {\tt PBCOR}.

\begin{figure}
\includegraphics[width=1.00\linewidth, angle=0]{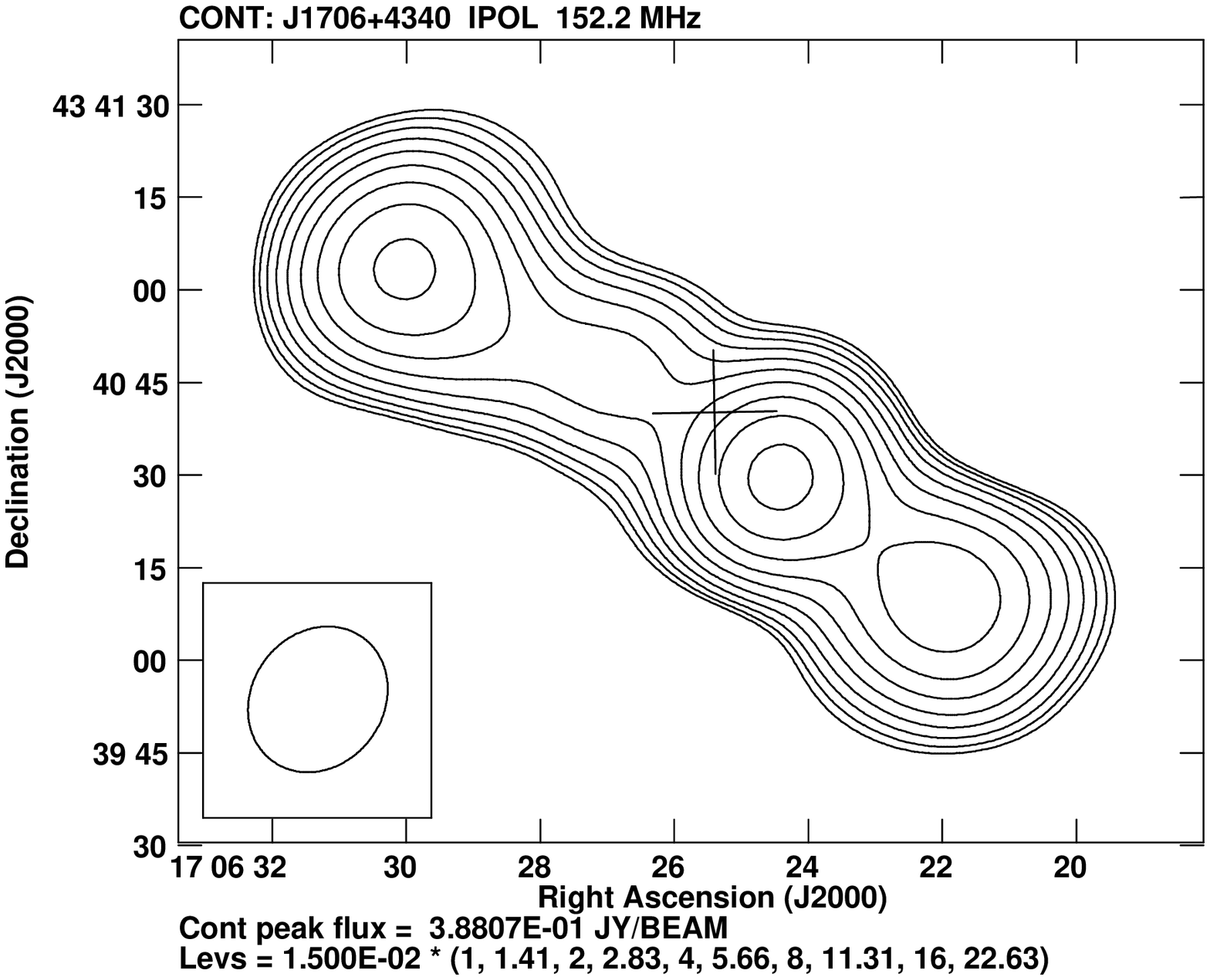}
\includegraphics[width=1.00\linewidth, angle=0]{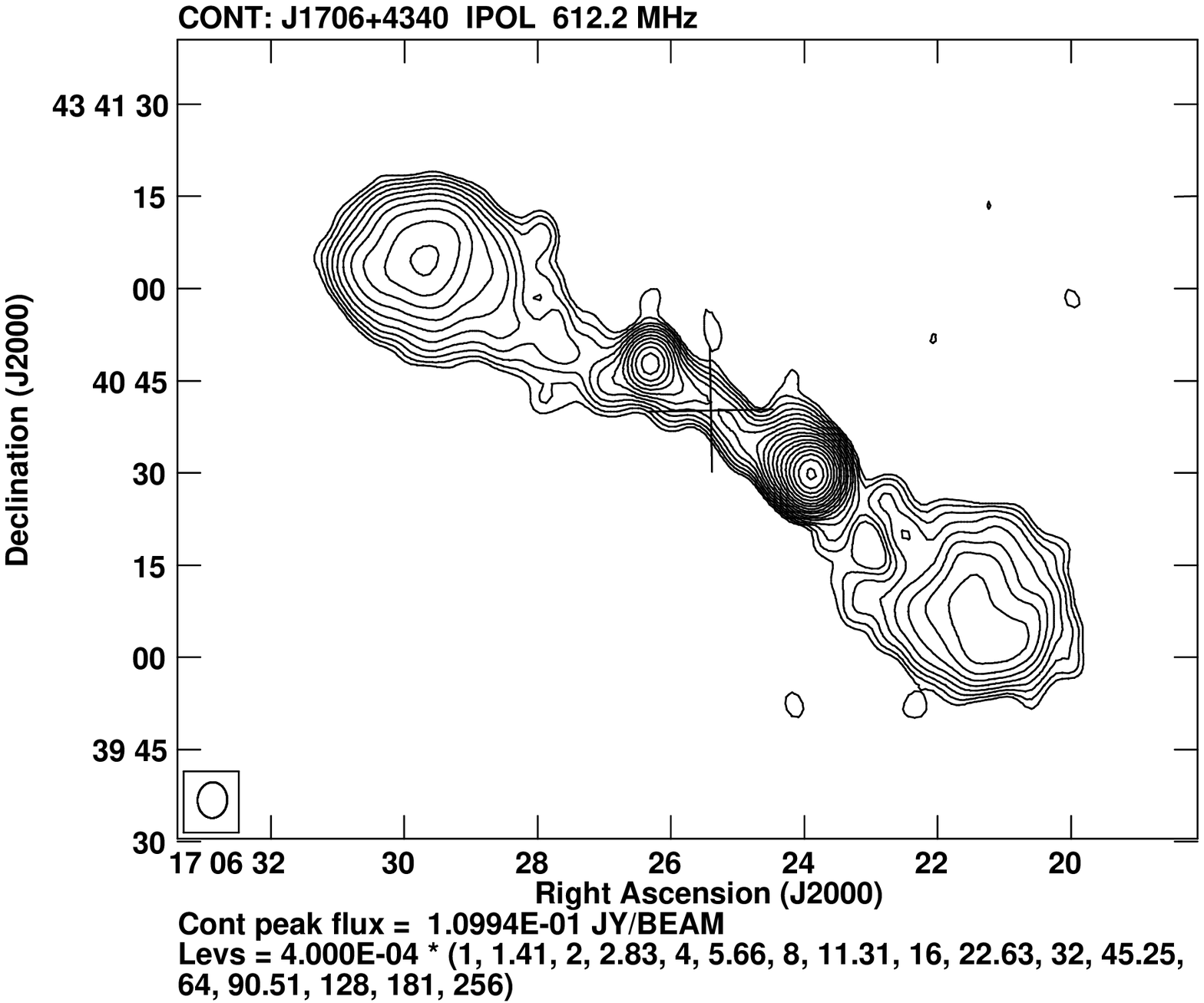}
\caption{GMRT images.
        Upper panel: 152.2\,MHz;
        Lower panel: 612.2\,MHz. 
Crosses in each panel indicate the position of the parent optical object.
The size of the beam is indicated by an ellipse in the bottom left corner of 
each panel.}
\label{fig:GMRT}
\end{figure}

The flux density calibration errors are assumed to be 5 per cent at 612.2 
and 8 per cent at 240.1 and at 152.2\,MHz. The flux density measurements and 
their errors are shown in Table\,2. The images at 152.2 and 612.2\,MHz are 
shown in Fig.\,\ref{fig:GMRT}. The quality of the 240.1\,MHz map is not 
satisfactory, therefore we do not present it here and use it only for the 
measurement of the total flux density of the source at this frequency.

\subsection{VLA observations and data reduction}
\label{VLA_data}

We observed J1706$+$4340 with the VLA at 1425 (conf. A and B), 4860 (conf. B 
and C), and 8460\,MHz (conf. B and C) between 2005\,April and\,2006 February. 
The exact dates of these observations along with the beam sizes are shown in 
Table\,1. They are marked with `VLA' followed by the letter A, B, or C -- it 
denotes the configuration -- in column (2) of that table and the letter `p'
in column (6). The length of each observation is given in the caption of 
Fig.\,\ref{fig:VLA}. 3C286 was used as an amplitude and J1707+4536 as a phase 
calibrator. Data reduction was carried out in {\sc aips} in a standard way. The
flux density measurements and their errors are shown in Table\,2. The images
were generated by {\sc aips} task {\tt IMAGR} and are shown in
Fig.\,\ref{fig:VLA}.

\subsection{Archival data}

\begin{figure}
\includegraphics[width=0.9\linewidth]{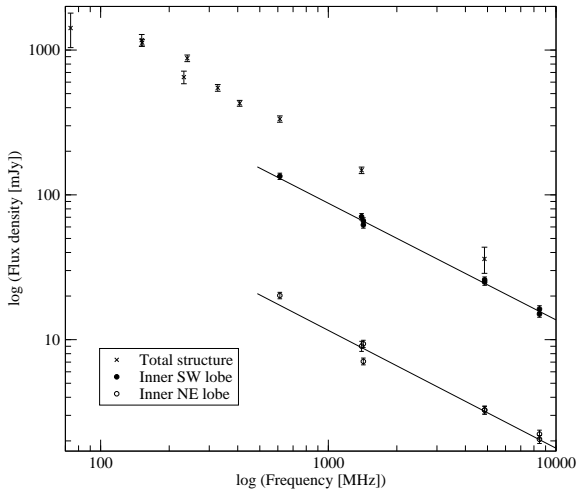}
\caption{Radio spectrum of the different parts of J1706$+$4340. 
Straight lines indicate power-law fit to the inner NE and SW lobes.}
\label{fig:spectr_obs}
\end{figure}

Apart from the dedicated observations itemized in 
subsections\,\ref{GMRT_data} and \ref{VLA_data}, we used publicly available 
images from the VLA Low-Frequency Sky Survey Redux at 74\,MHz (VLSSr; 
\citealt{Lane2014}), the Westerbork Northern Sky Survey at 327\,MHz (WENSS; 
\citealt{Rengelink1997}), FIRST, NRAO VLA Sky Survey at 1400\,MHz (NVSS; 
\citealt{Condon1998}), and the Green Bank 4850\,MHz northern-sky survey 
(GB6; \citealt{Gregory1996}). The flux densities obtained directly from the 
FITS maps (either produced for the purpose of this investigation or taken 
from the surveys) using {\sc aips} tasks {\tt TVSTAT} or {\tt JMFIT} (in the
case of point sources) are assumed to have the absolute flux calibration errors 
of 5 per cent. The flux densities taken from the literature are assumed to 
have the errors as determined in the corresponding references. In the case 
of the flux density values that were not measured directly from the FITS 
maps (e.g. those of the extended structure) but estimated by other means, we 
propagate the errors of the directly measured flux densities. In the case of 
an extended source, a noise term related to the size of its structure (the 
product of map noise and square root of the number of beams per structure) 
was added, and the overall error is a root mean square of the calibration 
and noise errors. The resulting flux densities (along with the errors) of 
our target, given in the \cite{Baars1977} scale are listed in Table\,2. In 
Fig.\,\ref{fig:spectr_obs}, we show the observational spectra of the whole 
source and of the inner north-eastern and south-western lobes. The spectra 
of the inner lobes are straight and flatter than the total source spectrum. 
Also, we obtained a combined spectrum of the inner lobes after adding the 
flux densities of both of them. Then we fitted a linear function to the data 
points in the form $\log[S (\rm{mJy})] = (-0.81\!\pm\!0.02)\times\log[\nu 
(\rm{MHz})] + (4.43\!\pm\!0.07)$. Finally, to obtain the flux densities of the 
outer lobes between 73 and 408\,MHz, we subtracted the predicted flux 
density of the inner lobes from the spectrum of the whole source. The 
remaining values at 612, 1400, and 4850\,MHz come from direct subtraction of 
the flux density figures of the inner structure from the flux density of the 
whole source.

\subsection{Overall radio structure}
\label{structure}

\begin{figure*}
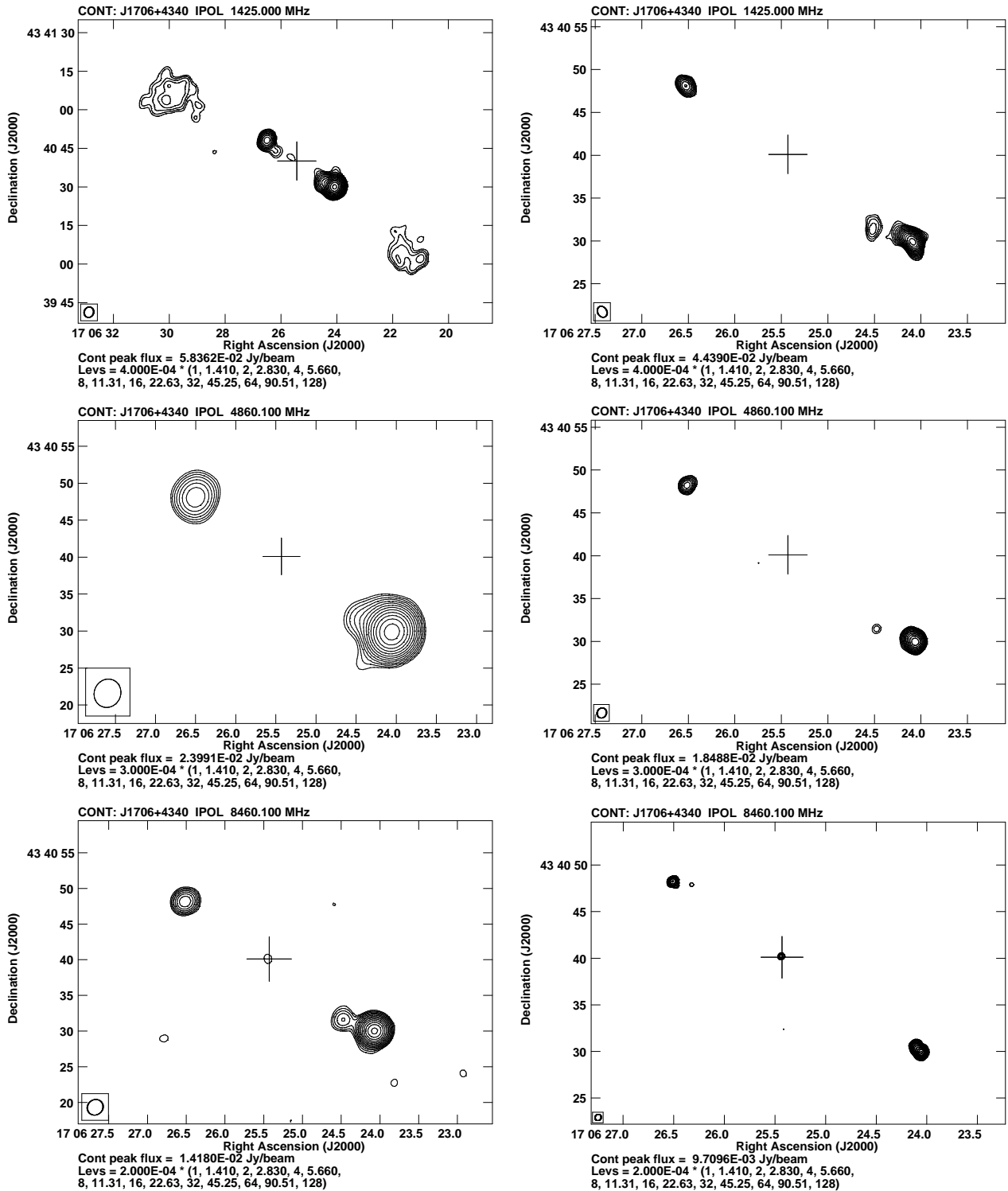

\includegraphics[width=0.48\linewidth, angle=0]{VLA.Bconf.Lband.PS}
\includegraphics[width=0.48\linewidth, angle=0]{VLA.Aconf.Lband.PS}
\includegraphics[width=0.48\linewidth, angle=0]{VLA.Cconf.Cband.PS}
\includegraphics[width=0.48\linewidth, angle=0]{VLA.Bconf.Cband.PS}
\includegraphics[width=0.48\linewidth, angle=0]{VLA.Cconf.Xband.PS}
\includegraphics[width=0.48\linewidth, angle=0]{VLA.Bconf.Xband.PS}
\caption{VLA images.
        Upper left: 1425\,MHz, B-conf., integration time on source --
        $10^{\rm m}50^{\rm s}$;
        Upper right: 1425\,MHz, A-conf., $9^{\rm m}45^{\rm s}$ on source;
        Middle left: 4860\,MHz, C-conf., $9^{\rm m}50^{\rm s}$ on source;
        Middle right: 4860\,MHz, B-conf., $18^{\rm m}55^{\rm s}$ on source;
        Lower left: 8460\,MHz, C-conf., $31^{\rm m}40^{\rm s}$ on source;
        Lower right: 8460\,MHz, B-conf., $28^{\rm m}45^{\rm s}$ on source.
Upper left panel encompasses the whole source, other panels -- only the inner
lobes. Crosses in each panel indicate the position of the parent optical
object. The size of the beam is indicated by an ellipse in the bottom left
corner of each panel.}
\label{fig:VLA}
\end{figure*}

The projected linear sizes of the inner and outer structures are 200 and 
750\,kpc, respectively. Based on the latter figure, J1706$+$4340 can be 
labelled a giant radio galaxy (GRG)\footnote{The minimum span of a GRG
is 1\,Mpc but for $H_0\!=\!50\,{\rm km\,s}^{-1}{\rm Mpc}^{-1}$. For $H_0$
adopted here, that lower limit should be downsized to 700\,kpc.}.
The 1400\,MHz luminosity of the inner and outer 
structures is $\rm \log(P_{inn})\!=\!25.88$\,W Hz$^{-1}$ and $\rm 
\log(P_{out})\!=\!25.85$\,W Hz$^{-1}$, respectively. This is a rare case of a 
DDRS whose inner double is as luminous as the outer one. There is a
characteristic brightness asymmetry in the lobes -- whereas the north-eastern 
outer lobe is only a little brighter than its opposite counterpart, the 
inner south-western lobe is much more luminous than the north-eastern one 
(cf. the fluxes in columns (3) and (4) of Table\,2). No hotspots are visible
in the outer lobes. This is particularly well pronounced in the 1400\,MHz VLA 
image (Fig.\,\ref{fig:VLA}, upper-left panel). The spectra of the outer 
lobes are very steep above 1400\,MHz, they are thus well imaged only with 
GMRT at 612.2\,MHz (Fig.\,\ref{fig:GMRT}, lower panel); at this frequency 
the lobes are still strong while the resolution of the instrument is the 
highest. Only the 8460\,MHz images reveal a very weak radio core at the
position of the optical counterpart. Its flux density is below 0.4\,mJy.

Among the six VLA images (Fig.\,\ref{fig:VLA}), the outer lobes can only be 
seen in the one acquired owing to the conf.\,B 1400\,MHz observation
(upper-left panel). In the case of conf.\,A observation at the same frequency,
the fraction of the `missing flux' due to diffuseness of the structures of the 
outer lobes must be close to unity, so that they are absent in that image. 
This is also the ultimate proof that the hotspots are missing there, otherwise 
they should have appeared there anyway due to their compactness. Because of 
diffuseness of structures and steepness of spectra as well as lack of 
hotspots, the outer lobes are not visible in the VLA images at 4860 and 
8460\,MHz.

\begin{table*}
\caption{Summary of the model parameters}
\begin{tabular}{lccc}
\hline
Parameter    & Symbol   & Outer lobes    & Inner lobes \\
\hspace{5mm}(1) & (2)  & (3)  & (4)   \\
\hline
{\bf Observed:} \\
Angular size of lobe ($\times 2$)           & LAS               & 120\,arcsec & 32\,arcsec \\
Length of lobe ($\times 2$)                 & $D$                 & 749 kpc      & 200 kpc \\
Aspect ratio of lobe                        & $R_{\rm T}$         & 2.15         & 10   \\
Radio spectrum; i.e. monochromatic          & $\alpha_{\rm\nu}$ \\
\hspace{5mm}luminosity at a number of observing \\
\hspace{5mm}frequencies; $i=1,2,3,$...     & $P_{\rm\nu,i}$      & Note(1)      & Note(2) \\
\vspace{1mm}
{\bf Set:} \\
Adiabatic index of jet material             & $\Gamma_{\rm j}$    & 4/3          & 4/3 \\
Adiabatic index of lobe material            & $\Gamma_{\rm\ell}$  & 5/3          & 4/3 \\
Adiabatic index of ambient medium           & $\Gamma_{\rm x}$    & 5/3          & 5/3 \\
Adiabatic index of magnetic field           & $\Gamma_{\rm B}$    & 4/3          & 4/3 \\
Minimum of initial electron Lorentz factor  & $\gamma_{\rm min}$  & 1            & 1   \\
Maximum of initial electron Lorentz factor  & $\gamma_{\rm max}$  & 10$^{7}$     & 10$^{7}$ \\
Core radius of ambient density distribution & $a_{0}$             & 2 kpc        & --  \\
Exponent of ambient density distribution    & $\beta$             & 1.5          & 1.5, 0 \\
Ratio of energy densities of thermal and \\
\hspace{5mm}relativistic electrons          & $k^{\prime}$        & 0, 1         & 0   \\
\vspace{1mm}
{\bf Free:} \\
Jet power                                   & $Q_{\rm j}$(W)  &\\
External density at core radius             & $\rho_{0}$(kg\,m$^{-3}$)   &\\
Exponent of initial power-law energy \\
\hspace{5mm}distribution of relativistic electrons  & $p$   &$=1+2\alpha_{\rm inj}$ \\
Source (lobe) age                           & $t$(Myr) &\\
\hline
\end{tabular}

\vspace{3mm}

{\em Notes.} (1) Relevant luminosities are calculated with flux densities 
shown in column (6) of Table\,2. (2) Relevant luminosities are calculated with 
flux densities shown in columns (3)+(4) of Table\,2.
\end{table*}

\section{Dynamical evolution analysis}

\subsection{Numerical code}
\label{num_cod}

Following the earlier work of \cite{Machalski2009, Machalski2010} and 
\cite{Machalski2011}, the dynamical analysis presented in this study has 
been carried out using {\sc dynage} code (\citealt{Machalski2007}) based on 
the analytic model for the evolution of FRII-type radio sources. It 
combines the pure dynamical model of \cite{KaiserAlexander1997} with the 
model of expected radio emission from a source under the influence of the 
energy loss processes elaborated by \cite{Kaiser1997} -- the KDA model.
For a given set of observables, this numerical code allows one to solve the 
inverse problem, i.e. to determine four free parameters of the KDA model: 
(1) the initial power-law energy distribution of the relativistic electrons 
$p$ related to the effective injection spectral index $\alpha_{\rm inj}$
(see Table 3), (2) the jet power $Q_{\rm j}$, (3) the density of the 
external gaseous medium near the radio core $\rho_{0}$, and (4) the age of 
the source's radio structure $t$.

A detailed description of how to apply the above code was given by 
\cite{Machalski2009}. Yet, it is worth mentioning here that determination of 
the values of these four parameters of the model is possible by a fit to the 
observational parameters of a source: its projected linear size $D$, the 
volume $V$, the radio power $P_{\nu}$ and the radio spectrum $\alpha_{\nu}$ 
that provides $(P_{\nu})_{\rm i}$ at a number of observing frequencies 
$i=1,\,2,\,3,...$. As in the KDA model, a cylindrical geometry of the 
source (its lobes) is assumed, $V\!=\!\uppi\,D^{3}/16\,R^{2}_{\rm T}$, where 
$R_{\rm T}$ is their axial ratio. Similarly, the minimum-energy condition 
for the ratio of the energy density of the magnetic field in the lobes to 
that of all particles is enforced by setting $r\!=\!(1+p)/[4(k^{\prime}+1)]$. 
This is justified by observations that the magnetized material in the lobes 
of FRII-type sources appears to be close to the minimum-energy conditions 
(cf. \citealt{Croston2005, Kataoka2005}). The values of the remaining free 
parameters of the model have to be assumed. These are the following: the core
radius $a_{0}$, the exponent $\beta$ describing the ambient density profile 
$\rho(d)\!=\!\rho_{0}(d/a_{0})^{-\beta}$, the adiabatic indices in equations
of state for the jet material, the magnetic field, the ambient medium, and the 
source (its lobes) as a whole, $\Gamma_{\rm j}$, $\Gamma_{\rm B}$, 
$\Gamma_{\rm x}$ and $\Gamma_{\rm\ell}$, respectively. The other free 
parameters to be assumed are the following: the Lorentz factors determining the
energy range of the relativistic electrons used in integration of their
power-law distribution $\gamma_{\rm i,min}$ and $\gamma_{\rm i,max}$, and the
ratio of the energy density of thermal particles to that of the relativistic 
electrons $k^{\prime}$.

The original KDA model assumes a constant and uninterrupted jet flow 
delivering relativistic gas to the expanding lobes throughout the whole 
lifetime of a source. However, the radio maps in Figs\,\ref{fig:GMRT} and 
\ref{fig:VLA} show  no evidence for active hotspots in the outer lobes.
This suggests that they are no longer supplied with energy by active jets.
Taking into account the presence of the inner lobes' structure, we assume
that the energy supply to the outer lobes ceased relatively recently in terms
of the total age of the source. Therefore, in order to analyse evolution of
this structure in the framework of the discussed scenario, we have to introduce
one more free parameter of the model -- the primary jet duration $t_{\rm j}$.

\begin{table}
\caption{Model and derivative parameters for the inner and the outer structure (lobes) of J1706$+$4340}
\begin{tabular}{lccccc}
\hline
                &\multicolumn{2}{c}{Inner structure} &  &\multicolumn{2}{c}{Outer structure}  \\ 
Parameters                  &$\beta$=1.5 &$\beta$=0  &  &$k^{\prime}$=0  &$k^{\prime}$=1 \\
\hspace{5mm}(1)             & (2)        & (3)       &    & (4)      & (5)  \\
\hline
{\bf Model:} \\
$\alpha_{\rm inj}$          & 0.554      & 0.544     &    & 0.526    & 0.54  \\
$Q_{\rm j}(\times 10^{37}$W)& 2.63       & 3.48      &    & 2.59     & 4.40  \\
$\rho_{0}(\times 10^{-22}$kg\,m$^{-3}$) & 5.43  & 0.075     &    & 18.6     & 22.3  \\
$t$(Myr)                    & --         & --        &    & 294      & 262   \\
$t_{\rm j}$(Myr)            & 11.8       & 12.3      &    & 275      & 244   \\
\\
{\bf Derivative:} \\
$v_{\rm h}/c(\times 10^{-3})$ & 23.7     & 15.9      &    & 2.86 & 3.20  \\
$\rho_{(D/2)}(\times 10^{-24}$kg\,m$^{-3}$)& 1.53  & 7.25   &    & 0.73     & 0.87  \\
$p_{\rm\ell}(\times 10^{-13}$N\,m$^{-2}$)  & 9.24  & 5.34   &  & 1.12     & 1.68  \\
$B_{\rm\ell}$(nT)            & 1.75       & 1.32     &    & 0.45     & 0.42  \\
\\
$\chi^{2}_{\rm red.}$       & 2.21       & 2.74      &    & 2.63     & 2.68  \\
\hline

\end{tabular}
\end{table}

\subsection{Application of the model and fitting procedure}

The analytical model described in Section\,\ref{num_cod} has been applied to
the observational data of the inner and outer lobes of J1706$+$4340.
The set of the model parameters with their values assumed for the given  
pair of lobes is shown in Table\,3.

\subsubsection{Inner lobes}
\label{inn_lobes}

As a first step, we apply the model to the observational data of the inner 
lobes because they are much better determined than those for the outer ones. 
Though the radio spectra for each of the inner lobes are derived, it should
be pointed out that their unusually high asymmetry in the observed 
luminosity ($\sim$1:8) poses a difficulty for the modelling. Indeed, such 
asymmetry induces significantly different values of the model parameters of 
individual lobes, like $Q_{\rm j}$, $\rho_{0}$, and $t$. In this context,
some self-consistent model solutions explaining the case by different 
ambient density profiles along the opposite jets were proposed by 
\cite{Machalski2009} and \cite{Machalski2011} but for the primary (outer) 
lobes in a sample of giant radio sources. A similar issue related to the 
secondary (inner) lobes of a known DDRS J1548$-$3216 was discussed by 
\cite{Machalski2010}, however in that case the relevant asymmetry was only 
about 1:2. The same aspect in two other DDRSs -- B1430+333 and B1843+620 -- 
was analysed and discussed by \cite{Brocksopp2011}. Here, we confine our 
analysis to the whole inner structure, i.e. omitting differences between the 
opposite lobes and using one half of the sum of the flux densities in 
columns (3) and (4) of Table\,2 to calculate the radio luminosities at a given 
observing frequency.

It has been pointed out by e.g. \cite{KaiserCotter2002}, \cite{Brocksopp2007}, 
\cite{Brocksopp2011}, \cite{Machalski2007}, and \cite{Machalski2010} that due
to a large number of 
free parameters, the applied analytical model is very flexible hence quite 
different sets of its parameters may fit the observational data -- 
especially the radio spectrum -- with a comparable level of significance. 
Since an independent observational constraint on the model parameters, e.g. 
on the ambient medium density provided by the X-ray measurements, is not 
available here, particularly for a density inside the outer lobes of 
J1706$+$4340, searching for a minimum of the jet energy delivered into the 
source, $(Q_{\rm j}\times t_{\rm j})_{\rm min}$ is included into our 
procedure.

\begin{figure}
\includegraphics[width=0.9\linewidth]{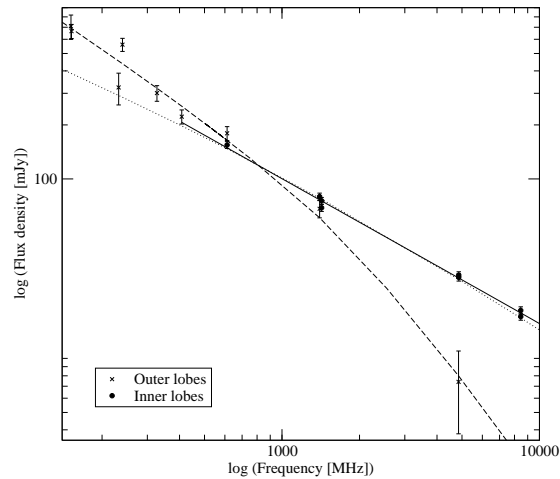}
\caption{Fitted spectra to the lobes. Straight line -- power-law fit to the
inner lobes. Dotted curve -- {\sc dynage} fit to the inner lobes. Dashed
curve -- {\sc dynage} fit to the outer lobes.}
\label{fig:spectr_model}
\end{figure}

We investigate two opposing models. In the first one, a power-law density 
distribution of the external gaseous medium with a standard exponent 
$\beta\!=\!1.5$ is assumed, while in the other one it is assumed that the 
inner lobes evolve into a pure uniform medium with $\beta\!=\!0$ inside a 
cocoon formed and inflated by the material of the primary jet flow. In both 
models, we assume that the lobes are filled only with magnetic fields and 
relativistic particles governed by a relativistic equation of state with 
$\Gamma_{\rm\ell}\!=\!4/3$. Columns (2) and (3) of Table\,4 show the fitting 
results for these two models, respectively. The relevant flux densities 
resulting from the first model are shown in column (5) of Table\,2 for a 
comparison with observations and Fig.\,\ref{fig:spectr_model} shows the 
model spectrum compared to the observed data points. The differences between 
various physical parameters resulting from these two models are further 
discussed in Sections\,\ref{age} and \ref{su_exp}.

\subsubsection{Outer lobes}
\label{out_lobes}

Contrary to the inner structure, the energy density of the outer hence older 
lobes may contain some contribution from thermal non-radiating particles, i.e.
$k^{\prime}\!\neq\!0$. We thus set $\Gamma_{\rm\ell}\!=\!5/3$ for these lobes. 
Note that when the jet activity is terminated, the adiabatic expansion of 
the primary lobes is expected to slow down and faster energy losses 
of the relativistic electrons are presumed. Therefore, two different phases 
of their dynamical evolution should be considered, the first one 
corresponding to the ongoing feeding of the lobes by the active jets (source 
ages $t\!\leq\!t_{\rm j}$) and the second one corresponding to the case when 
the jet flow has ceased (ages $t\!>\!t_{\rm j}$). Importantly, the evaluation
of the fading lobes' luminosities is done taking into account only those 
relativistic electrons that were injected into the lobes during the first 
phase of evolution, i.e. the ongoing jet activity. The integration over the 
injection time for the expected radio emission is performed following the 
prescription given by \cite{KaiserCotter2002} for relic radio sources.

As a first step, we search for the best model solution with $k^{\prime}\!=\!0$. 
The fitted values of the model parameters are shown in column (4) of Table\,4.
Note that the resultant values of $\alpha_{\rm inj}$ are almost identical 
in the best solutions for both the inner and outer lobes.
This is in an agreement with the main conclusion drawn by
\cite{Konar2013} based on a study of a sample of eight DDRSs (J1706$+$4340 
was not included there). Moreover, the values of $\alpha_{\rm inj}$ and 
$Q_{\rm j}$ we obtained follow the correlation between $\alpha_{\rm inj}$ 
and $Q_{\rm j}$ found by \citeauthor{Konar2013} -- see the bottom panel of
their fig.\,2 -- even though the method of determination of $\alpha_{\rm inj}$ 
and $Q_{\rm j}$ they used and the one we adopted here were different.
Those different approaches and their implications for the $\alpha_{\rm inj}$
and $Q_{\rm j}$ estimates are further discussed and commented in 
Section\,\ref{remarks}.

It should also be noted that the derived jet power during its primary activity
is comparable to its fitted value during the recurrent activity. The above 
result, obtained with the fitting procedure carried out independently for 
the outer and inner structures, may indicate that $Q_{\rm j}$ cannot vary 
significantly between the two active phases. However, the model of the inner 
lobes with $\beta\!=\!0$ admits slightly higher value of $Q_{\rm j}$ than that 
for the outer lobes. It is obvious that a contribution to the energy density 
of these lobes from possible thermal particles requires an increase of the 
jet power to explain their observed radio emission. Therefore, we 
investigate another model for the outer lobes with $k^{\prime}\!>\!0$, finding 
that $k^{\prime}\!=\!1$ already satisfied a condition $Q_{\rm j,out}\!\geq\! 
Q_{\rm j,inn}$. The relevant model parameters are shown in column (5) of 
Table\,4 and the flux densities resulting from the model with
$k^{\prime}\!=\!0$ -- in column (7) of Table\,2.
 
Except for the fitted values of the model parameters defined in 
Section\,\ref{num_cod}, some derivative physical parameters of the radio 
structures are also shown in Table\,4. These are the following: the 
longitudinal expansion velocity of the lobe given by differentiation of the 
formula for the jet length (cf. equations\,(4) and (5) in 
\citealt{KaiserAlexander1997}), the ambient density at the end of the lobe, 
$\rho(D/2)$, the pressure inside the lobe, $p_{\rm\ell}\!\equiv\! 
p_{\rm\ell}(t_{\rm j})$ and $p_{\rm\ell}(t)\!=\!p_{\rm\ell}(t_{\rm 
j})(t/t_{\rm j})^{-c_{1}}$ for the inner and the outer lobes, respectively, 
where $p_{\rm\ell}(t_{\rm j})$ corresponds to the cocoon's pressure given by 
equation\,20 in \cite{KaiserAlexander1997} and 
$c_{1}\!=\!6(\Gamma_{\rm\ell}-1)/(7+3\Gamma_{\rm\ell}-2\beta)$ (cf. 
\citealt{KaiserCotter2002}), and a mean value of the magnetic field 
strength, $B_{\rm\ell}\!=\!(2\mu_{0}u_{\rm B})^{1/2}$, where $u_{\rm 
B}\!=\!(p_{\rm\ell}\times r)/[(\Gamma_{\rm\ell}-1)(r+1)]$ is the magnetic 
field energy density.

Note that all the models reveal a similar agreement with the observational 
data. The last line in Table\,4 presents the $\chi^{2}$ values reduced by 
the degree of freedom. Although these values are rather high, they are 
overestimated by the large discrepancy between the flux densities measured 
on the Miyun Synthesis Radio Telescope (MSRT) and GMRT maps at $\sim$240\,MHz
used to determine the radio spectrum of the outer lobes
(cf. Fig.\,\ref{fig:spectr_model}), as well as likely underestimated flux
density at 1425\,MHz measured in the VLA A-conf. image in the spectrum of the
inner lobes.

\section{Discussion}

\subsection{Uncertain redshift}

Since the spectroscopic redshift of the host galaxy of J1706$+$4340 is not 
available, we use its photometric estimate -- it has been given in 
Section\,\ref{opt_data}. It is among the highest redshift values of radio 
galaxies suspected for a recurrent nuclear activity (cf. 
\citealt{Nandi2012}). Note that only B1843$+$620 galaxy ($z\!=\!0.519$) 
belongs to the group of well-studied DDRSs at redshift of $\sim\!0.5$. 
Therefore, the dynamical model parameters of J1706$+$4340 derived by the fit 
to its observational data, i.e. the energetics, ages, etc., can be less 
certain than those for sources with spectroscopic redshift available, and 
hence with more certain distance to the observer enabling a better 
determination of their size, luminosity, and radio spectrum. Taking into 
account possible overestimation of the adopted redshift, we performed the 
fitting procedure for the inner and the outer lobes adopting $z\!=\!0.3$.
Such hypothetical much lower value of redshift enables an investigation
of its influence on the fitted values of $\alpha_{\rm inj}$ and $Q_{\rm j}$.
Note that at this redshift both pairs of the lobes will appear less luminous 
and their span shorter compared to the respective values at $z\!=\!0.525$. 
In particular, the span of the inner/outer lobes will be $\sim\!140$ kpc and 
$\sim\!530$ kpc, respectively, placing J1706$+$4340 below the lower limit 
of the size of a GRG mentioned in Section\,\ref{structure}. Likewise a decreased
luminosity will imply a lower jet power adequate to provide the observed 
radio emission.
Indeed, we found the best-fitting solution with $\alpha_{\rm inj}\!=\!0.539$, 
$Q_{\rm j}\!=\!0.84\times 10^{37}$\,W and $t\!=\!15.8$\,Myr for the inner 
lobes modelled with $\beta\!=\!0$, as well as $\alpha_{\rm inj}\!=\!0.530$, 
$Q_{\rm j}\!=\!0.92\times 10^{37}$\,W, $t\!=\!340$\,Myr and $t-t_{\rm 
j}\!=\!28$\,Myr for the outer lobes assuming $k^{\prime}\!=\!1$. As 
expected, the best-fitting values of $\alpha_{\rm inj}$ and, especially, of 
$Q_{\rm j}$ diminish with decreasing redshift, thus being compatible with 
the $\alpha_{\rm inj}$ -- $Q_{\rm j}$ correlation noticed and discussed by 
\cite{Konar2013}.

\subsection{Remarks on the `spectral index--jet power' correlation and
their values during consecutive episodes of activity}
\label{remarks}

As noted in Section\,\ref{out_lobes}, our results obtained from the 
dynamical analysis follow the $\alpha_{\rm inj}$ -- $Q_{\rm j}$ correlation 
determined in a different way. \cite{Konar2013} used the synchrotron JP 
model (\citealt {Jaffe1973}) to fit the value of $\alpha_{\rm inj}$ and a 
frequency break $\nu_{\rm brJP}$ in radio spectra of the sources in their 
sample that allow determination of the spectral age of actually radiating 
relativistic electrons $\tau_{\rm syn}$. In order to estimate the value of 
$Q_{\rm j}$, they applied the relation equating the total energy deposited 
by the jet inside the volume of the lobe $V_{\rm\ell}$,
$(\Gamma_{\rm\ell}-1)\,Q_{\rm j}\,t_{\rm j}\approx p_{\rm\ell}V_{\rm\ell}$, 
admittedly identifying $t_{\rm j}$ with $\tau_{\rm syn}$. Their approach 
results in a steep slope of the investigated correlation. In fact, that 
slope flattens due to evidently higher jet powers if $\alpha_{\rm inj}$ and 
$Q_{\rm j}$ values are determined from the dynamical analysis. For example, 
\cite{Machalski2009} have already modelled most of the sources from the 
sample of \citeauthor{Konar2013} providing a comparison between spectral and 
dynamical $\alpha_{\rm inj}$ values, and finding $Q_{\rm j}$ values 
$\sim$(1.2--2) times higher than those in \citeauthor{Konar2013}. The 
latter result is easily explained by the fact that their approach accounts 
for the synchrotron losses only but not for the adiabatic expansion of the 
lobes.

However, it is worth emphasizing that \cite{Konar2013} have analysed 
and discussed an interesting problem of whether the injection spectral index 
and the jet power in the outer and the inner lobes of DDRSs are expected to 
differ or not. They pointed out that the principal factor determining 
$\alpha_{\rm inj}$ should be the strength of the jet-termination shock 
which, in turn, should depend on the external medium density $\rho_{\rm x}$. 
Since the inner lobes are immersed in the cocoon material of the outer lobes 
whose $\rho_{\rm x}$ is very likely different from the density of the 
primary environment surrounding the latter ones, one can expect a systematic 
difference between $\alpha_{\rm inj,inn}$ and $\alpha_{\rm inj,out}$, which 
is not observed. On the other hand, there is commonly expected that $Q_{\rm 
j}$ depends on the supermassive black hole (SMBH) mass, the accretion disc 
properties and accretion rate, and perhaps the SMBH spin (e.g. 
\citealt{Blandford1977}, \citealt{Tchekhovskoy2010}). The question why a 
correlation between the spectral index and the jet power -- the quantities 
that are ruled by quite different physical conditions inside the nucleus of 
a host galaxy and in the external environment -- exists, deserves further 
study.
 
\subsection{Age and intermittent activity}
\label{age}

The age solutions for the inner and outer lobes collected in Table\,4 
suggest that the last renewal of the jet activity took place about 12\,Myr 
ago, thus its present age is about $4\!-\!5$\ per cent of the age estimated 
for the outer lobes' structure. On the other hand, the quiescent period time 
emerging from the model fits, $t-t_{\rm j}$, of about $18\!-\!19$\,Myr
is not much larger than the age determined for the inner structure. 
However, it should be increased by the real time elapsed between the last 
jet material ejected by the AGN and the time it reached the head of the 
still-expanding lobe which is not taken into account in {\sc dynage} algorithm. 
When this is accounted for, the age proportions are very similar to those 
found in other already investigated DDRSs.

The first attempt to estimate the time-scales of the interruption of the jet 
flow was undertaken by \cite{Kaiserall2000}. They showed that this `elapsed' 
time is comparable to the age of inner lobes in their sample of five very 
large DDRSs including B1450$+$333 and B1834$+$620 further modelled by 
\cite{Brocksopp2011} in terms of their dynamical evolution. Also, 
\cite{Kaiserall2000} found comparable values for the ages of the inner lobes 
and the time lag $t-t_{\rm j}$, especially in the case of B1834$+$620. 
Another example of intensively studied DDRS is PKS\,B1545$-$321 
(J1548$-$3216). Its restarted jet activity was analysed and discussed by 
\cite{Saripalli2003}, \cite{Safouris2008}, and \cite{Machalski2010}. The 
inner lobes are deeply immersed into the extended cocoon of the primary 
radio emission implying a very short time-scale between cessation and renewal 
of the jet activity estimated as ($0.5\!-\!1$)\,Myr, while the fitted age of 
these lobes is $\sim(5\!-\!9)$\,Myr.

Duration of the quiescent phase can be also estimated from the curvature of 
the radio spectrum due to the energy losses caused by the synchrotron and 
inverse-Compton processes. In order to compare such estimate with that 
derived using {\sc dynage}, the spectral JP model has been fitted to 
the spectrum of the outer lobes of J1706+4340, and the value of $\nu_{\rm 
brJP}\!=\!1.8^{>40} _{-0.3}$\,GHz has been determined using {\sc synage} 
package (\citealt{Murgia1996}) adopting the injection spectral index 
$\alpha_{\rm inj}\!=\!0.54$, and the magnetic field strength 
$B_{\rm\ell}\!=\!0.42$\,nT, i.e. the values found using {\sc dynage}. It 
follows that duration of the period of quiescence is $27^{+2}_{-22}$\,Myr. 
As can be seen, the uncertainty of this period is very unsymmetrical; its 
upper limit is a little higher than ($18\!-\!19$)\,Myr estimated with the 
dynamical analysis, however rather expected as discussed above.

\subsection{Subsonic/supersonic expansion}
\label{su_exp}

Not only lacks J1706$+$4340 optical spectrum but there are also no X-ray 
observations available that, in principle, could provide some constraints on 
the model parameters including density, pressure, and kinetic temperature of 
the intergalactic medium surrounding the DDRS under investigation.
Nevertheless, a crude 
estimates of the ambient kinetic temperature and the sound speed in the 
ambient gaseous environment are possible using the derivative physical 
parameters shown in Table\,4. From the equation of state (for a perfect 
gas):

\[kT=\mu\,m_{\rm H}\,p/\rho\sim\mu\,m_{\rm H}\,p_{\rm h}/\rho_{\rm 
(D/2)}\hspace{3mm}{\rm and}\]

\[c_{\rm s}=[(\Gamma_{\rm x}\,kT)/(\mu\,m_{\rm H})]^{1/2},\]
 
\noindent
where the pressure of the jet's head (working surface of the bow shock) is 
taken as $p_{\rm h}\!=\! {\cal P}_{\rm h\ell}\,p_{\ell}$, and the ratio 
${\cal P}_{\rm h\ell}$ is adopted from \cite{Kaiser2000}. Assuming the mean 
atomic weight $\mu\!=\!0.62$, we find $kT\!=\!4.9$ keV and $c_{\rm 
s}\!=\!0.00374c$ for the outer lobes. Since the model with 
$k^{\prime}\!=\!0$ gives $v_{\rm h}/c\!=\!0.00286c$, the ratio $v_{\rm 
h}/c_{\rm s}\!=\!0.77$ supports a presumption based on a lack of hot spots 
and any trace of a bow shock on the radio maps, that the longitudinal 
expansion of these lobes is subsonic, i.e the Mach number $<1$. It is worth 
noting that the above ambient temperature is fully compatible with the X-ray 
temperatures in the samples of 20 powerful 3CR radio source
(\citealt{Belsole2007}), and 31 nearby clusters of galaxies 
(\citealt{Croston2008b}), fitted to the observed counts made with the {\em 
XMM--Newton} and {\em Chandra} observatories.

Estimation of the kinetic temperature inside the cocoon surrounding the 
inner lobes is problematic because their gaseous environment is likely 
modified by the jet flow during the earlier phase of activity. Thus, using 
the more conservative value of $p_{\ell}$ instead of $p_{\rm h}$ in 
calculation of $kT$ values from the two models investigated in 
Section\,\ref{inn_lobes} for these lobes, we find $kT\!=\!3.9$ keV, and $c_{\rm 
s}\!=\!0.0030c$ in the model with $\beta\!=\!1.5$, as well as $kT\!=\!0.5$ 
keV and $c_{\rm s}\!=\!0.0011c$ in the model with $\beta\!=\!0$. So, 
dividing the corresponding values of the advance speed of the inner lobes 
$v_{\rm h}/c$ (inserted in Table\,4) by the relevant sound speeds given 
above, we find the Mach number in the range 8--15. The value $kT\!<\!1$ keV 
estimated in the frame of the model with $\beta\!=\!0$ perhaps permits to 
distinguish the model with the uniform external density distribution, 
because such temperatures were found in the centres of haloes or clouds of 
X-ray emitting gas surrounding nearby radio galaxies (cf. 
\citealt{Allen2006, Lanz2015}). Also the $kT$ value, higher for the outer 
lobes than that for the inner ones, is in line with the electron density and 
temperature profiles presented in \cite{Croston2008a, Croston2008b}.

\section{Summary and conclusions}

\begin{enumerate}

\item In this article, we present a comprehensive study of a newly 
discovered DDRS\,J1706$+$4340. We carried out radio observations of this 
object with GMRT and the VLA at several frequencies ranging from 152
to 8460\,MHz and we present images resulting from these observations. The 
double--double structure is clearly visible in these images albeit only at 
frequencies lower than or equal to 1425\,MHz. The outer lobes are diffuse 
and show no traces of hotspots. It follows that they are in the coasting 
i.e. the final stage of their evolution.

\item We supplemented our data with those in the literature and constructed 
spectra of different parts of J1706$+$4340 covering more than two 
orders of magnitude wide frequency range: from 74 to 8460\,MHz. We modelled 
these spectra with {\sc dynage} code and determined the ages of both the outer
and inner lobes, as well as duration of the quiescent period. The resulting age 
of the outer lobes ($260\!-\!300$\,Myr) appears to be the oldest among 
relevant age estimates in other well-studied DDRSs, e.g. J0041$+$3224 
(\citealt{Saikia2006, Machalski2011}), B0925$+$420 (\citealt{Kaiserall2000, 
Brocksopp2007}), and those cited in Section\,\ref{age} where, unlike in 
J1706+4340, the majority of them still reveal either weak hotspots or a 
trace of bow shocks.

\item The frequency break, $\nu_{\rm br}$, determined from the fit of the JP 
model to the outer lobes' spectrum with {\sc synage}, together with the 
magnetic field strength, $B_{\rm\ell}$, found with {\sc dynage}, provide 
independent estimate of the quiescence period, fully compatible with that 
determined in the dynamical analysis. It is an order of magnitude shorter 
than the age of the outer lobes, whereas the age of the inner lobes is only 
$4\!-\!5$ per cent of the outer ones. All these results fit the well 
established paradigm ruling the DDRSs: they are restarted sources where the 
two pairs of lobes are pertinent to two consecutive episodes of activity
separated by a period of quiescence. Because of the appreciable steepness
of the spectra of the outer lobes above $\nu_{\rm br}$, combined with their
diffuseness, their apparent absence in the VLA images at 4860 and 8460\,MHz
may be considered explained.

\item The highly co-linear axes of both pairs of the lobes imply that the 
SMBH in the central AGN has a stable orientation. For the lack of noticeable 
distortions in the overall radio structure of J1706+4340, it is the spin of 
SMBH rather than e.g. an instability of the accretion disc that is likely to 
be responsible for the jet production. Therefore, it may be expected that 
the new jets, formed after disruption of their primary flow, should have the 
same power as the old jets. It is worth emphasizing that the above condition 
was almost fulfilled in {\sc dynage} solution shown in column (4) of 
Table\,4. However, we show that an admixture of non-radiating particles in 
the older outer lobes, parametrized by $k^{\prime}\!=\!1$, increases the 
energy requirement providing satisfactory surplus of the jet power.

\item We show how the pressure conditions in the lobes derived from {\sc dynage}
solutions allow a crude estimation of the kinetic temperatures of the 
gaseous environment surrounding the outer and the inner lobes as well as the 
relevant speeds of sound. We note that the temperature estimate around the 
outer lobes is fully compatible with the X-ray temperatures measured in the 
samples of powerful 3CR radio sources and nearby clusters of galaxies. We 
also find that the current longitudinal (very likely adiabatic) expansion of 
the outer lobes is already subsonic, thus compatible with presumption that 
they are in the coasting stage of their evolution.

On the other hand, the head expansion of the inner lobes is an order of 
magnitude faster than the outer ones, and the relevant Mach number is around 
10. Therefore, we conclude that these young lobes can substantially modify 
the surrounding matter by driving strong shocks and heating gaseous content 
of the old cocoon.

\item The extreme different values for the exponent of the external density
distribution along the inner lobes assumed in the fitting procedure with
{\sc dynage} though providing comparable ages and other model parameters
(except of $\rho_{0}$), imply clearly different kinetic temperatures of the
ambient medium, $kT\!>\!1$\,keV in the model with $\beta\!=\!1.5$ and
$kT\!<\!1$\,keV in that with $\beta\!=\!0$. It seems that the latter 
solution is preferred because such temperatures are typical for the X-ray 
emitting haloes detected around nearby radio galaxies.

\end{enumerate}

\section{Acknowledgements} We warmly thank A. Kurcz for performing the SED 
fitting with {\sc cigale}, as well as the referee for useful suggestions
helping to improve the paper. We acknowledge the access to {\sc synage} package 
provided by Dr Matteo Murgia. We thank the staff of GMRT that made these 
observations possible. GMRT is run by the National Centre for Radio 
Astrophysics of the Tata Institute of Fundamental Research. The National 
Radio Astronomy Observatory running the VLA is a facility of the National 
Science Foundation operated under cooperative agreement by Associated 
Universities, Inc. MJ and JM were supported by Polish NSC grant 
DEC-2013/09/B/ST9/00599.

\label{lastpage}

\end{document}